----- Begin Included Message -----

\documentstyle[12pt,fleqn]{article}

\newcommand{\bb}{\begin{eqnarray}}
\newcommand{\ee}{\end{eqnarray}}

\newcommand{\baa}{\begin{eqnarray*}}
\newcommand{\eaa}{\end{eqnarray*}}
\newcommand{\bebib}{\begin{thebibliography}}
\newcommand{\eb}{\end{thebibliography}}

\begin{document}

%

\title{The Free Field Representation of SU(3) Conformal Field Theory}
\author{Hovik D. Toomassian\thanks{Supported partly by a grant from
the Landau Institute Foundation.}\\ L.D.Landau Institute for Theoretical
Physics\thanks{Address after September 1 , 1992 : University of Southern
California , Physics Department , Los Angeles , California 90089 - 0484.},
\\Kosygina 2 , Moscow 117334 , Russia.}
\date{ November 1991}
\maketitle
\vspace*{6mm}

\begin{abstract}
The free field representation structure as well as some four point correlation
functions of the SU(3) conformal field theory are being considered.
\end{abstract}
\vspace{10mm}
\section{Introduction}

    Because of the role the two dimensional conformal field theories play in
 various fields of theoretical physics , these theories have been under
investi-
 gation for quite long time . One of the most effective ways of solving
conformal
 field theories, consists in reducing these theories to certain massless free
 field theories . The first models solved in this way were the minimal
models$^{1}$,
 solved completely on the sphere$^{2,3,4}$ . All the conformal blocks and
correlation
 functions were represented in Feigin - Fuchs integrals$^5$ and calculated
 explicitly . The operator algebra of the primary fields was derived then on
the
 basis of Polyakov's conformal bootstrap$^1$.

 Another example (still not the last) of free field realization is the SU(2)
conformal
field theory . It was initiated by Wakimoto's work$^6$ , where the Fock space
(or free field) representation of its chiral algebra  for general $k$  was
constructed.
The free field realization of the remaining ingredients of the theory ,
i.e.primary fields ,
screening operators and others , were given in$^7$ . Afterwards
the solution of the model was completed in $^{13,15}$ , where the correlation
functions
and the operator algebra were constructed . There is an alternative approach to
this
model$^8$ based on solving certain differential equations arising from the
symmetries
of the model . The results of these two solutions , of course , coincide.

A common belief exists now  that most of the conformal field theories admit
representations in terms of free field theories , though the Hilbert spaces of
the
conformal field theories appear to be only subspaces of the total Fock space of
the
free field theories$^{9,10,14}$ . A subsequent question arises  whether the
advances
with the minimal and SU(2) models could be consistently generalized to the
SU(N)
or any other conformal field theory . An essential step was made recently
towards this
generalization . The Wakimoto constructions were developed for general
Kac-Moody algebras$^{11}$
and general forms for the primary fields and screening operators were obtained
in
terms of a Heizenberg algebra generators (free fields) . Nevertheless the field
-
theoretical part of this generalization is , however , not transparent , yet .
Particularly
the integral representations of the conformal blocks are no more of Dotsenko -
Fateev
type$^{2,3}$ . Moreover , their analytic behaviours in their poles and zeroes
are unknown ,
which gives rise to formeadable difficulties in realizing the conformal
bootstrap$^1$ .

So , for these and other reasons it seems rather reasonable to start with the
free field
representation of the SU(3) conformal field theory on the sphere  suspecting
that it will
contain all the principle features of such theories .

This is what this paper is dealing with . Anyhow , it succeeds only in
performing the SU(2) parts of the SU(3) conformal
field theory.

It is organized as follows .
In section 2 the Wakimoto constructions , screening operators and the primary
fields are being introduced following the notations and terminology of $^{11}$
. A "Fock conjugated" primary operator is
introduced as well, to build the conformal blocks and correlation functions.

In section 3 the bosonized structures of
some SL(3) - modules are reviewed and certain four point correlation functions
are constructed and evaluated . The paper ends
with some concluding remarks . The holomorphic part of the theory is being
treated throughout the paper untill stated otherwise .

\section{Wakimoto constructions and primary fields}

To construct the free field realization of the SU(3) conformal field theory ,
one firstly needs the bosonized form of the SU(3) currents$^{11}$ :
\[E_{1}(z) = - :a_{11}^{*}a_{11}a_{11}^{*}: - \nu\varphi_{1}^{'}a_{11}^{*} - (2
- \nu^{2})\dot{a}_{11}^{*} + a_{22}a_{12}^{*} ;\]
\[F_{1}(z) = a_{11} + a_{12}a_{22}^{*} ;\]
\[H_{1}(z) = 2:a_{11}a_{11}^{*}: + :a_{12}a_{12}^{*}: - :a_{22}a_{22}^{*}: +
\nu\varphi_{1}^{'} ;\]
\[E_{2}(z) = - :a_{22}^{*}a_{22}a_{22}^{*}: + :a_{22}^{*}a_{11}a_{11}^{*}: -
:a_{22}^{*}a_{12}a_{12}^{*}: - \]
\[-\nu\varphi_{2}^{'}a_{22}^{*} - (3 - \nu^{2})\dot{a}_{22}^{*} -
a_{11}a_{12}^{*} ;\]
\begin{equation}
  F_{2}(z) =  a_{22}  ,
\end{equation}
\[H_{2}(z) =  2:a_{22}a_{22}^{*}: + :a_{12}a_{12}^{*}: - :a_{11}a_{11}^{*}: +
\nu\varphi_{2}^{'} ,\]
\[E_{3}(z) = - :a_{12}^{*}a_{12}a_{12}^{*}: - :a_{12}^{*}a_{22}a_{22}^{*}: -
:a_{12}^{*}a_{11}a_{11}^{*}: - \nu( \varphi_{1}^{'} + \varphi_{2}^{'}
)a_{12}^{*} - \]
\[ -(3 - \nu^{2})\dot{a}_{12}^{*} + (2 - \nu^{2})\dot{a}_{11}^{*}a_{22}^{*} +
:a_{11}a_{11}^{*}a_{11}^{*}a_{22}^{*}: + \nu\varphi_{1}^{'}a_{11}^{*}a_{22}^{*}
;\]
\[F_{3}(z) =   a_{12} ;\]
\[H_{3}(z) = 2:a_{12}a_{12}^{*}: + :a_{11}a_{11}^{*}: + :a_{22}a_{22}^{*}: +
\nu ( \varphi_{1}^{'} + \varphi_{2}^{'} ) \]
which are the generating functions for the Chevally generators \(E_{i}(n),
F_{i}(n), H_{i}(n)\)
\[E_{i}(z) = \sum_{n}E_{i}(n)z^{-n-1}\hspace{5mm} ;\hspace{5mm} E_{i}(n) =
\frac{1}{2i\pi}\oint{dz(z - w)^{n}E_{i}(z)} ;  \]
\begin{equation}
  F_{i}(z) = \sum_{n}F_{i}(n)z^{-n-1}\hspace{5mm} ; \hspace{5mm}F_{i}(n) =
\frac{1}{2i\pi}\oint{dz(z - w)^{n}F_{i}(z)} ;
\end{equation}
\[H_{i}(z) = \sum_{n}H_{i}(n)z^{-n-1}\hspace{5mm} ; \hspace{5mm}H_{i}(n) =
\frac{1}{2i\pi}\oint{dz(z - w)^{n}H_{i}(z)} .  \]
of the \(\hat{SL_k(3)}\) affine Kac - Moody algebra
\[[ E_{i}(n) , F_{j}(m) ] = \delta_{ij}H_{j}(n + m) + mk\delta_{m + n ,
0}\delta_{ij} ; \]
\[[ H_{i}(n) , E_{j}(m) ] = C_{ij}E_{j}(n + m) ; \]
\[[ H_{i}(n) , F_{j}(m) ] = - C_{ij}F_{j}(n + m) ;\]
\begin{equation}
  [ H_{i}(n) , H_{j}(m) ] = mkC_{ij}\delta_{n + m,0}  ;
\end{equation}
\[[ E_{i}(n) , E_{j}(m) ] = N_{ij}E_{i + j}(n + m); \]
\[[ F_{i}(n) , F_{j}(m) ] = - N_{ij}(n + m) .\]
Here and below \(N_{ij}=1 , \nu^{2}=k+\tilde{h} , 1\leq i,j\leq2\)
and $C_{ij}=2\delta_{ij}-\delta_{i,j-1}-\delta_{i,j+1}$ where k is the central
charge of the
$\hat{SL_{k}(3)}$ algebra , $C_{ij}$ is the Cartan matrix of the Lie algebra
SL(3) and $\tilde{h}$
is its dual Coxeter number.

In (2.1) the SU(3) currents are expressed in terms of free fields ( bosonic
ghosts ) $a_{ij}(z),a_{ij}^{*}(z)$ and $\varphi_{i}(z)$
$(1\leq i\leq j\leq 2)$ , where stress on $\varphi_{i}$ and dots on
$a_{ij}^{*}$ mean just derivatives of z . They
have the following two point functions and mode expansions
\[< \varphi_{i}^{'}(z) \varphi_{j}^{'}(w) > = \frac{C_{ij}}{(z - w)^{2}} ;\]
\begin{equation}
< a_{ij}(z) a_{pq}^{*}(w) > = - < a_{ij}^{*}(z) a_{pq}(w) > =
\frac{\delta_{ip}\delta_{jq}}{z - w} ;
\end{equation}
\[ \varphi_{i}^{'}(z) = \sum_{n}b_{i}(n)z^{-n-1} ;\hspace{2mm}a_{ij} =
\sum_{n}a_{ij}(n)z^{-n-1} ;\hspace{2mm}a_{ij}^{*} =
\sum_{n}a_{ij}^{*}(n)z^{-n-1} .\]
The Fourier components of these expansions realize a Heizenberg algebra
\[[ a_{ij}(n) , a_{pq}^{*}(m) ] = \delta_{ip}\delta_{jq}\delta_{n + m,0} ; \]
\begin{equation}
[b_{i}(n)  , b_{j}(m)] = n C_{ij} \delta_{n + m,0} ;
\end{equation}
\[[a , a] = [a^{*}, a^{*} ] = [a , b] = [a^{*} , b ] = 0  , \hspace{5mm}
\forall i, j, p, q, n, m .\]
The normal ordering in the bosonization formulas (2.1) means the common
prescription of
keeping the annihilation operators right to the creation ones, whenever they
are identified with respect to a vacuum
state in a representation of the Heizenberg algebra (2.5).

As is well known, the full symmetry algebra of the SU(N) conformal field
theories is a semiproduct of the $\hat{SL_k(N)}$
Kac-Moody and Virasoro algebras$^{12}$.The Virasoro algebra
\begin{equation}
[ L_{n} , L_{m} ] = ( n - m )L_{n+m} + \frac{c}{12} n(n^{2} - 1)\delta_{n + m ,
0}
\end{equation}
in these theories is realized by the Fourier components
$L_n$ of the stress-energy tensor T(z) , which is constructed as a normal
ordered inner product of the corresponding currents:
\begin{equation}
  T(z) = \frac{1}{2\nu^{2}} [\sum_{i,j=1,2} ( C_{ij}^{-1} : H_{i}H_{j} :  +  2
\delta_{ij} : E_{i}F_{j} : )]  ,
\end{equation}
where
\[T(z) = \sum_{n} \frac{L_n}{z^{n+2}} \hspace{5mm}     ,    \hspace{5mm}
L_{n} = \frac{1}{2i\pi}\oint{dz( z - w )^{ n + 1}} T(z) .\]
\vspace{5mm}
The central charge $c$ of the obtained Virasoro algebra can be caculated from
the two point function of the stress-
energy tensor (2.7) , using the operator product expansions between the SU(3)
currents
\vspace{10mm}
\[ E_{i}(z)F_{j}(w) = \frac{\delta_{ij}k}{(z - w)^{2}} + \frac{\delta_{ij}}{z -
w}H_{j}(w) + R.T. ;\]
\begin{equation}
   H_{i}(z)E_{j}(w) = \frac{C_{ij}}{z - w}E_{j}(w) + R.T. ;
\end{equation}
\[ H_{i}(z)F_{j}(w) = \frac{- C_{ij}}{z - w}F_{j}(w) + R.T. ;   \]
\[ H_{i}(z)H_{j}(w) = \frac{C_{ij}k}{(z - w)^{2}} + R.T. ,\]
where R.T. means terms which are regular as z \(\rightarrow\) w .
It yields to ,
\begin{equation}
\langle T(z)T(w) \rangle = \frac{4k}{k + 3} \frac{1}{(z - w)^{4}} =
\frac{c/2}{(z - w)^{4}} ,
\end{equation}
i.e.
\begin{equation}
 c = \frac{8k}{k + 3} = 8 - 24\alpha_{0}^{2}   ,   \alpha_{0}^{2} = (k +
3)^{-1} .
\end{equation}

Easy to obtain the bosonized form of the stress-energy
tensor T(z) , using the corresponding expressions for the currents (2.1) and
the two point functions (2.4) :
\begin{equation}
  T(z) = \sum_{i,j = 1,2}C_{ij}^{-1} ( \frac{1}{2} : \varphi_{i}^{'}
\varphi_{j}^{'} : + \frac{1}{\nu}\varphi_{j}^{"} ) +
\sum_{1\leq{i}\leq{j}\leq{2}} : \dot{a}_{ij}^{*}a_{ij} : .
\end{equation}
To pass to the field - theoretical content of the free field representation one
needs the bosonized forms of the primary fields$^1$
in the SU(3) conformal field theory.They were shown in$^{11}$ to have the
following general forms
\begin{equation}
\Phi(z) = \prod_{1\leq{i}\leq{j}\leq{rankg}}a_{ij}^{*^{d_{ij}}} V(\beta , z) ,
\end{equation}
where V($\beta$,z) is a vertex operator$^{11}$ and \(d_{ij}\) are positive
integer numbers . For the case of the SU(3) conformal field theory they admit
the following form
\begin{equation}
\Phi(z) = a_{11}^{*^{d_{11}}} a_{12}^{*^{d_{12}}} a_{22}^{*^{d_{22}}} exp (
\beta_{1}\varphi_{1} + \beta_{2}\varphi_{2} ) .
\end{equation}
In conformal field theories the primary fields serve as lowest weight vectors
of Verma modules over the Virasoro algebra ,
by their definition$^1$:
\begin{equation}
  T(z) \Phi(w) = \frac{\bigtriangleup}{(z - w)^{2}}\Phi(w)  +  \frac{1}{z -
w}\partial_{w}{\Phi(w)}  +  R.T.
\end{equation}
\[L_{0} = \Phi(w) = \bigtriangleup\Phi(w) ,\]
\begin{equation}
  L_{n} = o\hspace{5mm} ,\hspace{10mm}   if\hspace{5mm}  n\geq0 ;
\end{equation}
\[L_{n}\Phi(w) =  other\hspace{5mm} states \hspace{5mm}of\hspace{5mm}
the\hspace{5mm} module  \hspace{5mm} ,\hspace{5mm} if \hspace{5mm}n\leq0 .\]

To make the primary fields (2.12) serve as the lowest wieght vectors for the
SL(3) algebra modules also , they must be put to satisfy
\begin{equation}
  F_{i}(z)\Phi_{L.W.}(w) = R.T. , \hspace{20mm}   i = 1,2 .
\end{equation}
These conditions bring the primary fields (2.13) to the form of a "bare" vertex
operator
\begin{equation}
\Phi_{L.W.}(z) = V( \beta , z ) = exp( \vec{\beta} , \vec{\varphi} ) ,
\end{equation}
where two dimensional vectors \(\vec\beta=(\beta_{1},\beta_{2})\) and
\(\vec\varphi=(\varphi_{1} , \varphi_{2})\) are introduced for the sake of
having compact expressions.The operator product expansions of the primary field
(2.17) with the generators of the Cartan subalgebra are as follows
\begin{equation}
 H_{i}(z) \Phi_{L.W.}(w)   =   \frac{\nu ( \vec{\alpha_{i}} , \vec{\beta} )}{(z
- w)}\Phi_{L.W.}(w)  +  R.T. .
\end{equation}
Here \(\vec\alpha_{1}\),\(\vec\alpha_{2}\) are the simple roots of the SL(3)
Lie algebra ,
realized by the null components of the \(\hat{SL_k(3)}\) generators :
\begin{equation}
   [E_{i}(0) , F_{j}(0)] = \delta_{ij}H_{j}(0) , \hspace{5mm} [H_{i}(0) ,
F_{j}(0)] = - C_{ij}F_{j}(0) ,
\end{equation}
\[ [H_{i}(0) , E_{j}(0)] = C_{ij}E_{j}(0) ,  \hspace{3mm}   [H_{i}(0) ,
H_{j}(0)] = 0 .\]

All the remaining states of the \(\hat{SL_k(3)}\) modules are available , by
acting on the
lowest weight vector with the generators E$_{i}$(z) :
\begin{equation}
  E_{i}^{\Lambda_{i} + 1}(z) \Phi_{L.W.}^{(\lambda,\mu)}(w)  =  \frac{ - [
\Lambda_{i} + \nu(\vec\alpha_{i}, \vec\beta) ]}{(z -
w)}a_{ii}^{*^{\Lambda_{i}}}\Phi_{L.W.}^{(\lambda,\mu)}(w)  +  R.T..
\end{equation}

Easy to verify , that putting $\Lambda_{i}=-\nu(\vec\alpha_{i},\vec\beta)$ ,
finite dimensional irreducible
representations D$^{(\lambda,\mu)}$ of the SL(3) algebra are obtained with the
dimensions
\begin{equation}
  dim [D^{(\lambda , \mu)}]  =  \frac{1}{2} (\lambda + 1)(\mu + 1)(\lambda +
\mu + 2) ,
\end{equation}
where $\Lambda_{1}$ is denoted by $\lambda$ and $\Lambda_{2}$ is denoted by
$\mu$.This allows one to asign indices ($\lambda,\mu$)
\bb \left \{
\begin{array}{ll}
       \lambda =  \Lambda_{1} = - \nu ( \vec\alpha_{1} , \vec\beta ) = - \nu
\sum_{j=1,2} C_{1j}\beta_{j} , \\
       \mu = \Lambda_{2} = - \nu ( \vec\alpha_{2} , \vec\beta ) = - \nu
\sum_{j=1,2} C_{2j}\beta_{j} . \\
\end{array}
\right.  \ee
to the lowest weight vector (2.17), indicating its belonging to a certain
finite dimensional representation
D$^{(\lambda,\mu)}$ with the weight vector
$\vec\Lambda=(\lambda,\mu)$.Subsequently the conformal charges $\beta_{1}$ and
$\beta_{2}$ get quantized in "quantum numbers" of
the SL(3) algebra
\begin{equation}
 \beta_{i} = - \alpha_{0} \sum_{j=1,2} C_{ij}^{-1} \Lambda_{j} .
\end{equation}

Because the lowest weight vector (2.17) behaves analogous to a vector
\(\mid\alpha>\) of a bosonic Fock space F$_{\alpha}$ with respect to the
momentum p$_{i}^{14}$  \footnote{Recall the mode expansion of a
boson field $\varphi_{i}$(z) , and put [p$_{i}$,q$_{j}$]=C$_{ij}$.} ;
\begin{equation}
  p_{i} \mid\alpha> = \alpha_{i} \mid\alpha> \hspace{15mm} [ p_{i} ,
\Phi_{L.W.}^{(\lambda,\mu)}(z) ] = ( \vec\alpha_{i} , \vec\beta )
\Phi_{L.W.}^{(\lambda,\mu)}(z)  ;
\end{equation}
an identification is possible ,
\begin{equation}
  \Phi_{L.W.}^{(\lambda,\mu)}(0) \mid0>  =  \mid\vec\beta> , \hspace{15mm}
p_{i}\mid\vec\beta>  =  (\vec\alpha_{i} , \vec\beta) \mid\vec\beta> ;
\end{equation}
which allows one to refer to the above constructed modules as to "charged"
bosonic
Fock spaces F$_{\vec\beta,k}$ with conformal charge
$\vec\beta$=($\beta_{1},\beta_{2}$)$^9$ . These Fock spaces act as
representations of the Heizenberg algebra
\begin{equation}
  [ b_{i}(n) , b_{j}(m) ]  =  n C_{ij} \delta_{n + m, 0}
\end{equation}
with central elements b$_{i}$(0) , identified with the p$_{i}$ components of
the momentum $\vec{p}$ :
\begin{equation}
   [ b_{i}(0) , \Phi_{L.W.}^{(\lambda , \mu)}(z) ]  =  (\vec\alpha_{i} ,
\vec\beta) \Phi_{L.W.}^{(\lambda,\mu)}(z) .
\end{equation}

In consistency with the interpretitions above , the conformal blocks in free
field represented conformal field theories arise as expectation values in these
charged Fock spaces F$_{\vec{\beta},k}$ . So , to actually construct the
conformal blocks
, one needs the notion of the "Fock conjugated" lowest weight vector, to act as
the conjugate vacuum state$^{3,13}$ . On the other hand the appropriately
chosen conjugate vector allows one to avoid the redundent contour integrations
in the conformal
blocks$^{13}$ . This vector will be built with the same SL(3) quantum numbers
from a "Fock conjugated" primary field , represented in terms of the
contravariant Heizenberg algebra
\vspace{5mm}
\begin{equation}
   [  ^{\tau}b_{i}(n)  ,  ^{\tau}b_{j}(m) ]  =  - n C_{\tilde{h}-i ,
\tilde{h}-j} \delta_{m + n , 0 } ,
\end{equation}
\[ [  ^{\tau}a_{ij}(n)  ,  ^{\tau}a_{pq}^{*}(m) ]  =  - \delta_{ip} \delta_{jq}
\delta_{m + n , 0} , \]
as far as the conjugation attaches only the Fock space background of the
representation constructed above ,
with C$_{\tilde{h}-i,\tilde{h}-j}$=$ ^{\tau}$C$_{ij}=$C$_{ij}$ and $\tilde{h} =
3$ - being the dual Coxeter number of SU(3).

One can think then of the following expression for the general form of the
conjugate operator
\begin{equation}
   \tilde{\Phi}_{L.W.}^{(\lambda,\mu)}(z)  = ( ^{\tau}a_{11}^{*})^{q}(
^{\tau}a_{12}^{*})^{t}( ^{\tau}a_{22}^{*})^{s} exp[ ^{\tau}( \vec\beta
\vec\varphi ) ]  .
\end{equation}

Notice now that the same commutation relations (2.28) are satisfied if one
makes the following substitutions in the Heizenberg algebra (2.5)

\bb \left\{
\begin{array}{ll}
    b_{i}(n) \leftrightarrow - b_{\tilde{h}-i}(-n)   , \\

    a_{ij}(n) \leftrightarrow a_{ij}^{*}(-n)  . \\
\end{array}
\right.\ee
The isomorphism between these two sets of Heizenberg generators (2.28) and
(2.30) , allows one to put

\bb \left\{
\begin{array}{ll}
  ^{\tau}b_{i}(n) \equiv - b_{i}(-n)  ,\\

  ^{\tau}a_{ij}(n) \equiv a_{ij}(-n) .\\
\end{array}
\right. \ee

So one arrives then with the following form of the conjugate operator

\begin{equation}
  \tilde{\Phi}_{L.W.}^{(\lambda , \mu)}(z)  =  a_{11}^{q}a_{12}^{t}a_{22}^{s}
exp (\sum_{i = 1,2}\tilde\beta_{i} \varphi_{\tilde{h}-i}) .
\end{equation}
The parameters q,t,s, and $\tilde\beta_{i}$ are yet undefined.To
do this one must subject the conjugate operator (2.32) to the following
conditions . Firstly it must act as a lowest weight vector
\begin{equation}
   F_{i}(z) \tilde\Phi_{L.W.}^{(\lambda,\mu)}(w)  =  R.T. ,
\end{equation}
which implies s = 0.

Second , its SL(3) quantum numbers must coincide with those of the vector
(2.17) :
\begin{equation}
  H_{i}(z) \tilde\Phi_{L.W.}^{(\lambda,\mu)}(w)  =  H_{i}(z)
\Phi_{L.W.}^{(\lambda,\mu)}(w)  ;\hspace{15mm}   i = 1,2.
\end{equation}

The conditions (2.34) leave one with the following system of equations on the
parameters q , t , $\tilde\beta_{1} , \tilde\beta_{2}$ :
\bb \left\{
\begin{array}{ll}
       - 2q - t + \nu(2\tilde\beta_{2} - \tilde\beta_{1})  =  - \lambda   ,\\
         -q - t + \nu(2\tilde\beta_{1} - \tilde\beta_{2})  =  - \mu       . \\
\end{array}
\right. \ee

Third , because the $\tilde\Phi_{L.W.}^{(\lambda,\mu)}$(z) is thought to be a
primary field , all the coefficients at the singularities higher than the first
order in the operator product expansions
E$_{i}$(z)$\tilde\Phi_{L.W.}^{(\lambda,\mu)}$(w)
must be put to be zero :
\bb \left\{
\begin{array}{ll}
   - q(q - 1) - q(2 - \nu^{2}) + q\nu(2\tilde\beta_{2} - \tilde\beta_{1}) = 0
  ,\\
   - t(t - 1) - qt + \nu(\tilde\beta_{1} + \tilde\beta_{2}) - t(3 - \nu^{2}) =
0 .\\
\end{array}
\right.\ee

Solving the obtained equations on the parameters q , t , $\tilde\beta_{1}$ ,
$\tilde\beta_{2}$ (2.35) and (2.36) , four possible sets of these  parameters
are found . The last subjection to these parameters
\begin{equation}
  \bigtriangleup^{(\lambda , \mu)}  =  \tilde\bigtriangleup^{(\lambda , \mu )}
\end{equation}
leaves one with two of the four solutions :
\begin{equation}
q = 0 , t = 0 , \tilde\beta_{1} = \beta_{2} , \tilde\beta_{2} = \beta_{1}
\end{equation}
and
\begin{equation}
q = 0 , t = - \eta + \nu(\tilde\beta_{1} + \tilde\beta_{2}) , \tilde\beta_{1} =
\eta\alpha_{0} - \beta_{1} , \tilde\beta_{2} = \eta\alpha_{0} - \beta_{2}  ,
\end{equation}
where $\eta$ = - k - 1 .

The first one coincides with the lowest weight vector (2.17).So one can refer
to the second set , as to the parameters of the conjugate operator , for it
satisfies all the subjected conditions (2.33) - (2.37) . To rewrite (2.39) in a
more compact
way, it's convinient to introduce the following two vectors
$\vec\alpha_{0}$=($\alpha_{0},\alpha_{0}$) and
$\Theta$=$\vec\alpha_{1}$+$\vec\alpha_{2}$ . Then
\begin{equation}
       t = \eta + ( \vec\Theta , \vec\Lambda ) \hspace{20mm}, \hspace{20mm}
\vec{\tilde\beta} = \eta\vec\alpha_{0} - \vec\beta
\end{equation}

\[  \tilde\beta_{i}  = \eta\alpha_{0}  +  \alpha_{0}\sum_{j=1,2}
C_{\tilde{h}-i,\tilde{h}-j}^{-1} \Lambda_{j} . \]

Finally for the conjugate lowest weight vector , the following free field
representation is constructed
\begin{equation}
  \tilde\Phi_{L.W.}^{(\lambda , \mu)}(z)  =  a_{12}^{\eta + (\vec\Theta ,
\vec\Lambda)} exp (\sum_{i = 1,2}\tilde\beta_{i}\varphi_{\tilde{h}-i}) .
\end{equation}

The existance of such a lowest weight vector is interpreted in charged Fock
space terms , as a manifestation of an isomorphism between the dual Fock space
F$_{\vec\beta,k}^{*}$ and the Fock space F$_{\eta\vec\alpha_{0}-\vec\beta,k}$ ,
if one puts
\begin{equation}
  ^{\tau}b_{i}(0) \equiv \eta(\alpha_{0})_{i} - b_{\tilde{h}-i}(0) ,
\end{equation}

Another important ingredient of the theory is obtained , when one puts
$\vec\Lambda$=(0,0) and $\vec\beta$=(0,0) :
\begin{equation}
  \tilde\Phi_{L.W.}^{(\lambda,\mu)}(z)
\hspace{10mm}\rightarrow\hspace{10mm}\tilde{\bf1}(z) =
a_{12}^{\eta}exp[\sum_{i =
1,2}\eta(\alpha_{0})_{i} \varphi_{\tilde{h}-i}] .
\end{equation}

It is called the conjugate identity operator , because it commutes with the
current algebra (2.1) and has a null conformal dimension ,
like the identity operator $\Phi_{L.W.}^{(0,0)}$(z) = {\bf 1}.

The existance of such an operator is a consequence of the conformal charge
asymmetric structure of the Fock space expectation values.

The last necessary objects for constructing the conformal blocks in a free
field represented theory , are the so called screening operators.They have the
following form in the case of the SU(3) theory :

\[ J_{1}^{\pm}(z) = a_{11}^{n_{\pm}} exp (\alpha_{\pm} \varphi_{1}) ; \]
\begin{equation}
   J_{2}^{\pm}(z) = (a_{22} + a_{11}^{*}a_{12})^{n_{\pm}} exp (\alpha_{\pm}
\varphi_{2}) .
\end{equation}
where n$_{-}$= - (k+3) , n$_{+}$=1 , $\alpha_{+}$=  - $\alpha_{-}^{-1}$ =
$\alpha_{0}$ = ($\sqrt{k+3}$)$^{-1}$.

Actually , for the purposes pursued in this paper , two of them  J$_{1}^{+}$
and  J$_{2}^{+}$ will do quite well.

The screening operators have peculiar operator product expansions with the
current algebra generators (2.1).The only non-vanishing operator product
expansions ( i.e. expansions with singularities ) have the following forms of
total derivatives
\[ E_{i}(z) J_{j}^{+}(w)  =  \delta_{ij} (k + 3)
\partial_{w}(\frac{exp(\alpha_{0}\varphi_{j})}{z - w}) ,\hspace{16mm}  i = 1,2
\]
\begin{equation}
   E_{i + j}(z) J_{i}^{+}(w)  =  (i - j)(k + 3)
\partial_{w}(\frac{a_{jj}^{*}exp(\alpha_{0}\varphi_{i})}{z - w}) .
\end{equation}
which allows one to use them under the contour integrals effectively , while
building the conformal blocks .

It's worth noting as well , that this screening operators have their own ,
independent of the conformal blocks, interpretitions in algebraic
contents$^{11,14}$ . And the possability of using them effectively in conformal
blocks originates from the
BRST - cohomological structure of the free field representations of the
conformal field theories$^{9,10,11,14}$.

So much for the preliminaries in dealing with the correlation functions and
conformal blocks in the SU(3) conformal theory represented in free fields.

\section{SL(3) modules and correlation functions}
\setcounter{equation}{0}

The prescription of building nonvanishing N - point functions in conformal
field theories$^{2,3}$ requires the conformal charges $\vec\beta$ to satusfy
certain constraints,called the neutrality conditions .
In the case of the SU(3) conformal field theory they have the following form
\begin{equation}
 \sum_{a = 1}^{N - 1}\vec\beta_{a} + \vec{\tilde\beta_{N}} =
\eta\vec{\alpha_{0}}  ,
\end{equation}
if one needs to calculate an N - point function of the type
\begin{equation}
  \langle \Phi_{1}^{(\lambda_{1} , \mu_{1})}(z_{1} , \bar{z}_{1})  \ldots
\Phi_{N - 1}^{(\lambda_{N-1} , \mu_{N-1})}(z_{N-1} , \bar{z_{N-1}})
\tilde\Phi_{N}^{(\lambda_{N} , \mu_{N})}(z_{N} ,
\bar{z_{N}}) \rangle .
\end{equation}
To take into consideration the screening currents (2.45) ,
one modifies the equations (3.1)$^{2,3}$ :
\begin{equation}
 \sum_{a = 1}^{N - 1}\vec{\beta}_{a}+\vec{\tilde\beta_{N}}+\vec{n}\alpha_{0}=
\eta\vec{\alpha}_{0}     ,
\end{equation}
which leads in its turn , to the following values for the numbers of screening
currents needed
\begin{equation}
  n_{i} = \sum_{j = 1,2} [ \sum_{a = 1}^{N - 1}C_{ij}^{-1}\Lambda_{j}^{a} -
C_{ij}^{-1}\Lambda_{\tilde{h}-J}^{N} ] .
\end{equation}
\vspace{10mm}
Here $\vec{n}=(n_{1} , n_{2})$ is a formal vector , $n_{1}$ being the number of
screening currents $J_{1}^{+}$ and $n_{2}$ - the number of $J_{2}^{+}$ .
\footnote{$\vec{n}$ can be interpreted as the weight vector of
a ($\vec{n_{1}} , \vec{n_{2}}$) representation of the SL$_{q}$(3).}

For the ($a , a^{*}$) - part of the N - point function (3.2) the charge
conservation conditions are subjected in the following way

\bb \left\{
\begin{array}{ll}
      N (a_{12}) + n_{2} - N (a_{12}^{*}) = \eta \\
      N (a_{22}) + n_{2} - N (a_{22}^{*}) = 0 \\
      N (a_{11}) + n_{1} - n_{2} - N(a_{11}^{*}) = 0   \\
\end{array}
\right. \ee
where N($a$) is the numbers of $a$ - fields of certain type in the correlator
(3.2) .
These conditions are equivalent to the claim for the N - point function (3.2)
to be SL(3) invariant , i.e. to have the SL(3) charges conserved , because the
SL(3) part of the representation of the symmetry algebra of the
theory is spanned by  the field operators $a_{ij}(z)$ and $a_{ij}^{*}(z)$ . The
conserved SL(3) charges are the third component of isospin $T_{z}$ and the
hypercharge $Y$ . They are connected with the Cartan basis generators
$H_{1}$ and $H_{2}$ in the following way
\bb \left\{
\begin{array}{ll}
      T_{z} = \frac{1}{2} H_{2} \\
      Y = \frac{2}{3} H_{1} + \frac{1}{2} H_{2} \\
\end{array}
\right.\ee

In these quantities equations (3.5) are equivalent to more transparent
equations
\bb \left\{
\begin{array}{ll}
     \sum_{a = 1}^{N} T_{z}^{a} = 0 \\
     \sum_{a = 1}^{n} Y^{a} = 0 \\
\end{array}
\right.\ee
which are very convinient when correlation functions are being considered
\footnote{Here the same notations for the generators $T_{z}$ and $Y$ in (3.6)
and their eiginvalues in (3.7) are used , which can't cause misunderstanding in
what follows .}.
As a matter of fact all the information for having nonvanishing correlation
functions is encoded in the equations (3.3) and (3.7) .

For a four point correlation function one gets from the equations (3.4) :
\bb \left\{
\begin{array}{ll}
       3n_{1} = \sum_{a = 1}^{3}(2\lambda_{a} + \mu_{a}) - (2\mu_{4} +
\lambda_{4}) \\
       3n_{2} = \sum_{a = 1}^{3}(2\mu_{a} + \lambda_{a}) - (2\lambda_{4} +
\mu_{4}) \\
\end{array}
\right.\ee

The consideration of these equations shows that most of the four point
correlation functions (3.2) with insertions of both screening currents
$J_{1}^{+}$ and $J_{2}^{+}$ possess conformal block functions , the structure
of zeroes and poles of which are yet unclear and they don't admit reductions to
the Dotsenko - Fateev type hypergeometric functions  (at least
naively) . The main problem is , that the functional equation which is
satisfied by the normalization integrals of the conformal blocks (see Appendix
A in $^3$) turn out to be "less informative" in this case . Mainly ,
due to the fact of non - commutativity ( in a quantum group sense ) of the
screening currents $J_{1}^{+}$ and $J_{2}^{+}$ . Nevertheless , it seems that
the Dotsenko - Fateev proceedure for calculating the normalization
integrals ( though perhaps being not the only possible way ) must be appliable
to all the generic cases of conformal field theories with proper modifications
, supported by the quantum group structure of the space of conformal
 blocks . In this paper the consideration of four point correlation functions
is restricted to the ones,reducible to Dotsenko - Fateev type integrals . These
are mainly the four point correlation functions which have only
 one type of screening currents inserted . They represent, in a certain sense,
the SU(2) parts of the SU(3) conformal field theory .

For such kind of four point correlation functions the neutrality conditions
(3.8) can be rewritten in the following ways :
\[   \vec{n} = ( n_{1} , 0 ) \]
\bb \left\{
\begin{array}{ll}
    \sum_{a = 1}^{3}\mu_{a} = \lambda_{4} - n_{1} \\
    \sum_{a = 1}^{3}\lambda_{a} = \mu_{4} + 2n_{1} \\
\end{array}\right.\ee

\[ \vec{n} = ( 0 , n_{2} ) \]
\bb \left\{
\begin{array}{ll}
     \sum_{a = 1}^{3}\lambda_{a} = \mu_{4} - n_{2} \\
     \sum_{a = 1}^{3}\mu_{a} = \lambda_{4} + 2n_{2} \\
\end{array}
\right.\ee

A sort of duality in ($\lambda , \mu$) and ($n_{1} , n_{2}$) present in (3.9)
and (3.10)  is
worth noting .

The most convinient choice of the states from SL(3) multiplets , which will
stand in four point correlators satisfying the SL(3) charge conservation (3.7)
, appears to be the following one
\begin{equation}
 \langle \Phi_{H.W.}^{(\lambda_{1} , \mu_{1}}(z_{1} , \bar{z_{1}})
\Phi_{H.W.}^{(\lambda_{2} , \mu_{2})}(z_{2} , \bar{z_{2}}) \Phi_{T_{z}^{3} ,
Y^{3}}^{(\lambda_{3} , \mu_{3})}(z_{3} , \bar{z_{3}})
\tilde\Phi_{L.W.}^{(\lambda_{4} , \mu_{4})}(z_{4} ,
\bar{z_{4}})\rangle .
\end{equation}
Here H.W ( L.W. ) means the highest ( lowest ) weight vector of a particular
SL(3) module ( $\lambda , \mu$ ) . The choice of ( $T_{z}^{3} , Y^{3}$ ) is
fixed then by solving consistently the equations (3.9) and (3.10) with (3.7) .
The solutions  yield to the following choices for ( $T_{z}^{3} , Y^{3}$ ) in
correspondence with (3.9) and (3.10)

 \[ \vec{n} = ( n_{1} , 0 ) \]

\bb \left\{
\begin{array}{ll}
   T_{z}^{3} = \frac{\lambda_{3}}{2} - n_{1} = T_{z , H.W.}^{3} - n_{1} \\
   Y^{3} = \frac{1}{3}(2\mu_{3} + \lambda_{3}) = Y_{H.W.}^{3}             \\
\end{array}
\right.\ee
\vspace{10mm}
 \[ \vec{n} = ( 0 , n_{2} ) \]

\bb \left\{\begin{array}{ll}
   T_{z}^{3} = \frac{\lambda_{3}}{2} + \frac{n_{2}}{2} = T_{z , H.W.}^{3} +
\frac{n_{2}}{2} \\
   Y^{3} = \frac{1}{3} (2\mu_{3} + \lambda_{3}) - n_{2} = Y_{H.W.}^{3} - n_{2}
\\
\end{array}
\right.\ee

In the equations (3.12) and (3.13) the relations between the "physical" and
"Cartan" quantum numbers are used for the H.W and L.W. states of a
representation ($\lambda , \mu$) :
\[ H.W.s. = ( T_{z} , Y ) = ( \frac{1}{2}\lambda , \frac{1}{3}(2\mu + \lambda)
) \]
\begin{equation}
 L.W.s. = ( T_{z} , Y ) = ( -\frac{1}{2}\mu , - \frac{1}{3}(2\lambda + \mu) )
\end{equation}

To be more concrete in evaluating the four point correlation functions (3.11)
and for the sake of simplicity as well , let's consider the cases when only
representations of
the ($\lambda , 0$) and/or ($0 , \mu$) types are inserted where possible.

For such insertions the neutrality conditions (3.9) and (3.10) together with
(3.12) and (3.13) lead to the following conformal blocks for the four point
correlation function\footnote{Here and below the convencional notations for the
conformal
blocks and other ingredients of the theory set up in$^{2,3}$ are used.} (3.11)
:

\[ I^{( n_{1} , 0 )}(z_{1},z_{2},z_{3},z_{4}) \sim
\int_{s_{1}}dv_{1}\int_{s_{2}}dv_{2}\ldots\int_{s_{n_{1}}}dv_{n_{1}}  \times \]
\[\hspace{15mm}\times\langle
\Phi_{H.W.}^{(\lambda,0)}(z_{1})\Phi_{H.W.}^{(n_{1},0)}(z_{2})\Phi_{T_{z}^{3},Y^{3}}^{(n_{1},0)}(z_{3})\tilde\Phi_{L.W.}^{(n_{1},\lambda)}(z_{4}) \times \]
\begin{equation}
  \hspace{15mm}\times J_{1}^{+}(v_{1}) J_{1}^{+}(v_{2})\ldots
J_{1}^{+}(v_{n_{1}}) \rangle
\end{equation}

\[ I^{( 0 , n_{2} )}(z_{1},z_{2},z_{3},z_{4}) \sim
\int_{c_{1}}du_{1}\int_{c_{2}}du_{2}\ldots\int_{c_{n_{2}}}du_{n_{2}}  \times \]
\[\hspace{15mm}\times\langle \Phi_{H.W.}^{(0 , \mu)}(z_{1})\Phi_{H.W.}^{(0 ,
n_{2})}(z_{2})\Phi_{T_{z}^{3},Y^{3}}^{(0 ,
n_{2})}(z_{3})\tilde\Phi_{L.W.}^{(\mu , n_{2})}(z_{4}) \times \]
\begin{equation}
\hspace{15mm}\times J_{2}^{+}(u_{1}) J_{2}^{+}(u_{2})\ldots
J_{2}^{+}(u_{n_{2}}) \rangle
\end{equation}

Let's now have a look on how the modules of ($\lambda , 0$) and (0 , $\mu$)
types are constructed .

The root diagram of the SL(3) algebra (2.19) , Fig.1. , induces the following
weight diagrams for the representations ($\lambda$ , 0) and (0 , $\mu$) ,
Fig.2. and Fig.3. .

In general , the states in these modules can be represented in the following
way
\vspace{5mm}
\begin{equation}
 \Phi_{i,j}^{(\lambda , 0)}(z) = : E_{2}^{i} E_{1}^{\lambda - j} V^{(\lambda ,
0)} : (z)
\end{equation}
\begin{equation}
 \Phi_{i,j}^{(0 , \mu)}(z) = : E_{2}^{i} E_{3}^{\mu - j} V^{(0 , \mu)} : (z)
\end{equation}
with the corresponding weights

\[  H_{1}(z) \Phi_{i,j}^{(\lambda , 0)}(w) = \frac{ - ( 2j - \lambda + i )}{z -
w}\Phi_{i,j}^{(\lambda , 0)}(w) + R.T. \]
\begin{equation}
    H_{2}(z) \Phi_{i,j}^{(\lambda , 0)}(w) = \frac{2i - \lambda +j}{z -
w}\Phi_{i,j}^{(\lambda , 0)}(w) + R.T. \\
\end{equation}
\vspace{5mm}
\[    H_{1}(z) \Phi_{i,j}^{(0 , \mu)}(w) = \frac{\mu - i -j}{z -
w}\Phi_{i,j}^{(0 , \mu)}(w) + R.T. \]
\begin{equation}
      H_{2}(z) \Phi_{i,j}^{(0 , \mu)}(w) = \frac{2i - j}{z - w}\Phi_{i,j}^{(0 ,
\mu)}(w) + R.T.
\end{equation}
Or equivalently

\bb \left\{\begin{array}{ll}
    T_{z} = \frac{1}{2} ( - \lambda + j + 2i ) \\
    Y = \frac{1}{3} ( \lambda - 3j )\\
\end{array}
\right.\ee
in ($\lambda , 0$) and

\bb \left\{\begin{array}{ll}
    T_{Z} = - \frac{1}{2} ( j - 2i ) \\
    Y = \frac{1}{3} ( 2\mu - 3j )\\
\end{array}
\right.\ee
in ($0 , \mu$).

Naivly one needs the explicit bosonized forms of the r.h.s. of (3.17) and
(3.18) to evaluate the conformal blocks (3.15) and (3.16).

In fact it turns quite sufficient  having the bosonized states from some of the
simple modules like $\lambda=1,2,3$ and $\mu=1,2,3$ to generalize then the
results for the conformal blocks of the form (3.15) and (3.16) containing
arbitrary $\lambda$
and $\mu$.

In the Fig.4. the corresponding weight diagrams are represented for the modules
(1 , 0) , (2 , 0) , (3 , 0) and for their duals on the Fig.5. . Missing the
rather tidious calculations let's have just the bosonized forms of the statesin
the modules of
Fig.4. :
\vspace{10mm}
\[ \Phi_{L.W.}^{(1,0)} = \Phi_{0,1}^{(1,0)} = V^{(1,0)} \]
\begin{equation}
   \Phi_{0,0}^{(1,0)} = a_{11}^{*} V^{(1,0)}
\end{equation}
\[ \Phi_{H.W.}^{(1,0)} = \Phi_{1,0}^{(1,0)} = ( a_{12}^{*} -
a_{22}^{*}a_{11}^{*} ) V^{(1,0)} \]

\vspace{5mm}
\[ \Phi_{L.W.}^{(2,0)} = \Phi_{0,2}^{(2,0)} =
V^{(2,0)}\hspace{10mm}\Phi_{1,0}^{(2,0)} = ( a_{22}^{*}a_{11}^{*^{2}} -
a_{12}^{*}a_{11}^{*} ) V^{(2,0)} \]
\begin{equation}
   \Phi_{0,1}^{(2,0)} = a_{11}^{*} V^{(2,0)}\hspace{15mm}\Phi_{1,1}^{(2,0)} = (
a_{12}^{*} - a_{11}^{*}a_{22}^{*} ) V^{(2,0)}
\end{equation}
\[ \Phi_{0,0}^{(2,0)} = a_{11}^{*^{2}}
V^{(2,0)}\hspace{15mm}\Phi_{H.W.}^{(2,0)} = \Phi_{2,0}^{(2,0)} = (
a_{22}^{*}a_{11}^{*^{2}} - 2a_{12}^{*}a_{11}^{*}a_{22}^{*} + a_{12}^{*^{2}} )
V^{(2,0)} \]
\vspace{5mm}

\[ \Phi_{L.W.}^{(3,0)} = \Phi_{0,3}^{(3,0)} =
V^{(3,0)}\hspace{10mm}\Phi_{1,0}^{(3,0)} = ( a_{22}^{*}a_{11}^{*^{3}} -
a_{12}^{*}a_{11}^{*^{2}} ) V^{(3,0)}\]
\[ \Phi_{0,2}^{(3,0)} = a_{11}^{*} V^{(3,0)}\hspace{21mm}\Phi_{2.0}^{(3,0)} = (
a_{22}^{*^{2}}a_{11}^{*^{3}} - 2a_{22}^{*}a_{12}^{*}a_{11}^{*^{2}} +
a_{12}^{*^{2}}a_{11}^{*} ) V^{(3,0)}\]
\[ \Phi_{0,1}^{(3,0)} = a_{11}^{*^{2}}V^{(3,0)}\hspace{20mm}\Phi_{1,1}^{(3,0)}
= ( a_{11}^{*^{2}}a_{22}^{*} - a_{12}^{*}a_{11}^{*} ) V^{(3,0)} \]
\begin{equation}
   \Phi_{0,0}^{(3,0)} = a_{11}^{*^{3}}V^{(3,0)}\hspace{20mm}\Phi_{1,2}^{(3,0)}
= (a_{11}^{*}a_{22}^{*} - a_{12}^{*} ) V^{(3,0)}
\end{equation}
\[ \Phi_{H.W.}^{(3,0)} = \Phi_{3,0}^{(3,0)} = ( a_{11}^{*^{3}}a_{22}^{*^{3}} -
a_{12}^{*}a_{11}^{*^{2}}a_{22}^{*^{2}} + 3a_{11}^{*}a_{22}^{*}a_{12}^{*^{2}} -
a_{12}^{*^{3}} ) V^{(3,0)}\]
\[ \Phi_{2,1}^{(3,)}  = ( 6a_{11}^{*}a_{22}^{*}a_{12}^{*} - 3a_{12}^{*^{2}} -
a_{11}^{*^{2}}a_{22}^{*^{2}} ) V^{(3,0)}\]
\vspace{5mm}

For the states in the dual modules (0 , 1) , (0 , 2) , (0 ,3) , Fig.5., one
arrives with :

\[ \Phi_{L.W.}^{(0,1)} = \Phi_{0,1}^{(0,1)} = V^{(0,1)}\]
\begin{equation}
   \Phi_{1,1}^{(0,1)} = a_{22}^{*} V^{(0,1)}
\end{equation}
\[ \Phi_{H.W.}^{(0,1)} = \Phi_{0,0}^{(0,1)} = a_{12}^{*} V^{(0,1)}\]
\vspace{5mm}
\[ \Phi_{L.W.}^{(0,2)} = \Phi_{0,2}^{(0,2)} =
V^{(0,2)}\hspace{10mm}\Phi_{1,1}^{(0,2)} = a_{12}^{*}a_{22}^{*}V^{(0,2)}\]
\begin{equation}
   \Phi_{1,2}^{(0,2)} = a_{22}^{*}V^{(0,2)}\hspace{15mm}\Phi_{0,1}^{(0,2)} =
a_{12}^{*}V^{(0,2)}
\end{equation}
\[ \Phi_{2,2}^{(0,2)} = a_{22}^{*^{2}}V^{(0,2)}\hspace{15mm}\Phi_{H.W.}^{(0,2)}
= \Phi_{0,0}^{(0,2)} = a_{12}^{*^{2}}V^{(0,2)}\]
\vspace{5mm}
\[ \Phi_{L.W.}^{(0,3)} = \Phi_{0,3}^{(0,3)} =
V^{(0,3)}\hspace{10mm}\Phi_{2,2}^{(0,3)} = a_{12}^{*}a_{22}^{*^{2}}V^{(0,3)}\]
\[ \Phi_{1,3}^{(0,3)} = a_{12}^{*}V^{(0,3)}\hspace{15mm}\Phi_{1,1}^{(0,3)} =
a_{12}^{*^{2}}a_{22}^{*}V^{(0,3)}\]
\begin{equation}
   \Phi_{2,3}^{(0,3)} = a_{22}^{*^{2}}V^{(0,3)}\hspace{15mm}\Phi_{H.W.}^{(0,3)}
= \Phi_{0,0}^{(0,3)} = a_{12}^{*^{3}}V^{(0,3)}
\end{equation}
\[ \Phi_{3,3}^{(0,3)} = a_{22}^{*^{3}}V^{(0,3)}\hspace{15mm}\Phi_{0,1}^{(0,3)}
= a_{12}^{*^{2}}V^{(0,3)}\]
\[ \Phi_{0,2}^{(0,3)} = a_{12}^{*}V^{(0,3)}\hspace{15mm}\Phi_{1,2}^{(0,3)} =
a_{22}^{*}a_{12}^{*^{2}}V^{(0,3)}\]
\vspace{5mm}

Now let's examine the states appearing in the conformal blocks (3.15) and
(3.16) from the modules ($\lambda , 0$) and ($0 , \mu$) .
As was mentioned in the sect. 2 , for the sake of simplicity it's convinient to
to keep the "Fock conjugated" representations
($\tilde{\lambda , \mu}$) represented in the conformal blocks by the lowest
weight states (2.41) . Anyhow , there's no obstacle in constructing  "Fock
conjugated" modules ($\tilde{\lambda , 0}$) and ($\tilde{0 , \mu}$) ,
acting explicitly in the same way as in (3.17) and (3.18) . All the states in
these "Fock conjugated" modules have the same corresponding weights (3.19) and
(3.20) .
The only difference is that in the case of the "Fock conjugated" modules null
states occur in the
places where numerical zeroes while biulding the modules ($\lambda , 0$) and
($0 , \mu$) appear (e.g.
E$_{1}^{\lambda+1}(z)\tilde{\Phi}_{0,0}^{(\lambda,0)}$(w) = null state ).

The states standing at the point (z$_{3}$ , $\bar{z_{3}}$) from the modules
(n$_{1} , 0$) and ($0 , n_{2}$) in (3.15) and (3.16) have the following
bosonized forms . From the equations (3.12) and (3.13)
one learns that these are the states with ($-n_{1}/{2} ,
$Y$_{H.W.}$) from (n$_{1} , 0$) in the case of (3.12) and (n$_{2}/{2} ,
$Y$_{H.W.}-$n$_{2}$) from ( $0 , $n$_{2}$ ) in the case of (3.13) . Taking into
consideration (3.21) and (3.22) one arrives with the following states , ( $j =
0
, i = 0 $) for the
case of (n$_{1} , 0$) and ($j =$ n$_{2}$ , $i =$ n$_{2}$) for the case of (0 ,
n$_{2}$) .

Examining the explicit bosonized forms for the states in the modules ($\lambda
, 0$) (eqs. 3.23 - 3.25) and ($0 , \mu$) (eqs. 3.26 - 3.28) as well as the
Fig.4. and Fig.5. one can notice that in general the states
$\Phi_{0,0}^{(\lambda,0)}(z)$
( $\Phi_{\mu,\mu}^{(0,\mu)}(z)$ ) are $\lambda$ order ($\mu$ order ) monomials
of the fields $a_{11}^{*}(z)$ ( $a_{22}^{*}(z)$ ) . This leads to another
statement , that the only terms from the ($a ,a^{*}$)
part of the highest weight states of ($\lambda , 0$)
modules having  nonzero contributions in the correlator (3.15) are the ones
containing the fields $a_{12}^{*}(z)$ in the corresponding $\lambda$th power .
In the case of the correlator (3.16) this observation turns to be obvious ,
as far as the highest weight states in the modules ($0 , \mu$) are just $\mu$th
order monomials of $a_{12}^{*}(z)$ .

Concluding these preliminary observations let's write down the exact forms of
the holomorphic conformal blocks (3.15) and (3.16) for general values of
n$_{1}$ = p and n$_{2}$ = q in the projective group gauge $z_{1}=0 ; z_{2}=z ;
z_{3}=1 ; $
$z_{4}\rightarrow\infty$ as usual :

\vspace{5mm}

\[ I^{(p , 0)}(z) =
\int_{s_{1}}dv_{1}\int_{s_{2}}dv_{2}\ldots\int_{s_{p}}dv_{p}\times\]
\vspace{5mm}
\[ \hspace{5mm}\times\langle V^{(\lambda , 0)}(0)V^{(p , 0)}(z)V^{(p ,
0)}(1)\tilde{V}^{(p , \lambda)}(\infty)V_{1}^{+}(v_{1})V_{1}^{+}(v_{2})\ldots
V_{1}^{+}(v_{p})\rangle\]
\vspace{5mm}
\begin{equation}
   \hspace{5mm}\times\langle
a_{12}^{*^{\lambda}}(0)a_{12}^{*^{p}}(z)a_{11}^{*^{p}}(1)a_{12}^{\eta + p +
\lambda}(\infty) a_{11}(v_{1})a_{11}(v_{2})\ldots a_{11}(v_{n_{1}})\rangle
\end{equation}
\vspace{6mm}
\[ I^{(0 , q)}(z) =
\int_{c_{1}}du_{1}\int_{c_{2}}du_{2}\ldots\int_{c_{q}}du_{q}\times\]
\vspace{5mm}
\[ \hspace{5mm}\times\langle V^{(0 , \mu)}(0)V^{(0 , q)}(z)V^{(0 ,
q)}(1)\tilde{V}^{(\mu , q)}(\infty)V_{2}^{+}(u_{1})V_{2}^{+}(u_{2})\ldots
V_{2}^{+}(u_{q})\rangle\]
\vspace{5mm}
\begin{equation}
   \hspace{5mm}\times\langle
a_{12}^{*^{\mu}}(0)a_{12}^{*^{q}}(z)a_{22}^{*^{q}}(1)a_{12}^{\eta + q +
\mu}(\infty)a_{22}(u_{1})a_{22}(u_{2})\ldots a_{22}(u_{q})\rangle
\end{equation}
\vspace{10mm}

Evaluating the constructed conformal blocks (3.29) and (3.30) with the help of
the corresponding two point functions (2.4) , one gets finally the following
expressions of them :

\[ I^{(p , 0)}(z) = z^{\alpha_{12}} (1 - z)^{\alpha_{23}} \int_{s_{1}}dv_{1}
\int_{s_{2}}dv_{2}\ldots \int_{s_{p}}dv_{p}\times\]
\vspace{5mm}
\begin{equation}
   \hspace{15mm}\times \prod_{i = 1}^{p}v_{i}^{a} (1 - v_{i})^{b - 1}(z -
v_{i})^{c} \prod_{i < j}(v_{i} - v_{j})^{2\rho}
\end{equation}
\vspace{10mm}
\[ I^{(0 , q)}(z) = z^{\alpha_{12}^{'}}(1 -
z)^{\alpha_{23}^{'}}\int_{c_{1}}du_{1}\int_{c_{2}}du_{2}\ldots\int_{c_{q}}du_{q}\times\]
\vspace{5mm}
\begin{equation}
   \hspace{15mm}\times\prod_{i = 1}^{q} u_{i}^{a^{'}}(1 - u_{i})^{b^{'} - 1}(z
- u_{i})^{c^{'}}\prod_{i < j}(u_{i} - u_{j})^{2\rho}
\end{equation}

Here ;
\[\alpha_{12} = \vec\beta^{(\lambda , 0)} C \vec\beta^{(p ,
0)}\hspace{5mm};\hspace{5mm}\alpha_{12}^{'} = \vec\beta^{(0 , \mu)} C
\vec\beta^{(0 , q)}\]
\[\alpha_{23} = \vec\beta^{(p , 0)} C \vec\beta^{(p ,
0)}\hspace{5mm};\hspace{5mm}\alpha_{23}^{'} = \vec\beta^{(0 , q)} C
\vec\beta^{(0 ,q)}\]
\[a = \vec\beta^{(\lambda , 0)} C \vec{e_{1}}\hspace{5mm};\hspace{5mm}a^{'} =
\vec\beta^{(0 , \mu)} C \vec{e_{2}}\]
\begin{equation}
  b = c = \vec\beta^{(p , 0)} C \vec{e_{1}}\hspace{5mm};\hspace{5mm}b^{'} =
c^{'} = \vec\beta^{(0 , q)} C \vec{e_{2}}
\end{equation}
\[\vec{e_{1}} = (\alpha_{0} , 0)\]
\[\vec{e_{2}} = (0 , \alpha_{0})\]
\[2\rho = \vec{e_{1}} C \vec{e_{2}} = 2\alpha_{0}^{2} \]
\vspace{10mm}

The integrals (3.31) and (3.32) are obviously of Dotsenko - Fateev type, the
technique of handling which as well as their structures of zeroes and poles are
worked out in$^{2,3}$ . Due to the commutativity of the fields $a_{11}^{*}(z)$,
$a_{22}^{*}(z)$ and $a_{12}^{*}(z)$ the integrals (3.31) and (3.32) appear to
be simpler than in the case of the SU(2) conformal field theory ( comp.$^{15}$
) .  In referrences$^{2,3}$ a convinient basis of linearly independant
conformal blocks was
performed identifying the basis elements by a system of independent contour
configurations , which asigns the conformal blocks a "monodromy" index $k$ in
the following way :
\vspace{15mm}
\[ I_{k}^{(p , 0)}(a,b,c;\rho;z) = \int_{1}^{\infty}dv_{1} \ldots\int_{1}^{v_{p
- k - 1}}dv_{p - k}\int_{0}^{z}dv_{p - k + 1}\ldots\int_{0}^{v_{p -
1}}dv_{p}\times\]

\[\hspace{15mm}\times\prod_{i = 1}^{p}v_{i}^{a}\prod_{i = 1}^{p - 1}(v_{i} -
1)^{b - 1}(v_{i} - z)^{c}\prod_{i = p - k + 1}^{p}(1 - v_{i})^{b - 1}(z -
v_{i})^{c}\]

\begin{equation}
  \hspace{15mm}\times\prod_{i < j}(v_{i} - v_{j})^{2\rho}
\end{equation}
\vspace{6mm}
\[ I_{k}^{(0 , q)}(a^{'},b^{'},c^{'};\rho;z) = \int_{1}^{\infty}du_{1}
\ldots\int_{1}^{u_{q - k - 1}}du_{q - k}\int_{0}^{z}du_{q - k +
1}\ldots\int_{0}^{u_{q - 1}}du_{p}\times\]

\[\hspace{15mm}\times\prod_{i = 1}^{q}u_{i}^{a^{'}}\prod_{i = 1}^{q - 1}(u_{i}
- 1)^{b^{'} - 1}(u_{i} - z)^{c^{'}}\prod_{i = q - k + 1}^{q}(1 - u_{i})^{b^{'}
- 1}(z - u_{i})^{c^{'}}\]

\begin{equation}
  \hspace{15mm}\times\prod_{i < j}(u_{i} - u_{j})^{2\rho}
\end{equation}
\vspace{10mm}

The overall factors z$^{\alpha_{12}}$(1 - z)$^{\alpha_{23}}$
are suppressed as far as they are common to all the integrals and can be
restored in the final expressions . Afterwards the consideration of the
conformal blocks (3.34) and (3.35) will be restricted to one of them ,
particularly to the first one , keeping
in mind that the analogous simulations are appliable to the second one , when a
, b , c , and $\alpha$'s are substituted by the stressed ones and p by q.
The choice of the basis (3.34) in the space of the conformal block functions
corresponds to the expansion of the four point function over the s - channel
partial waves$^{2,3,15}$ .

The basis functions I$_{k}^{(n_{1},0)}(a,b,c;\rho;$z$)$ can be represented in
the following way :
\vspace{10mm}
\begin{equation}
  I_{k}^{(p , 0)}(a,b,c;\rho;z) = N_{k}^{(p ,
0)}(a,b,c;\rho)F_{k}(a,b,c;\rho;z) = N_{k}^{(p , 0)} z^{\gamma_{k}} f_{k}(z)
\end{equation}
\vspace{10mm}
where $f_{k}$(z) is analytic at z = 0 ,$f_{k}$(0) = 1 and $\gamma_{k} = (k -
1)(1 + a + c + (k - 2)\rho)$ . $N_{k}^{(p , 0)}$(a,b,c;$\rho$;z) is the so -
called normalization integral , having the following form ($^3$ , Appendix A )
:
\vspace{10mm}
\[ N_{k}^{(p , 0)} = \int_{0}^{1}dt_{1}\int_{0}^{t_{1}}dt_{2} \ldots
\int_{0}^{t_{p - k -1}}dt_{p - k}\int_{0}^{1}ds_{1}\int_{0}^{s_{1}}ds_{2}
\ldots \int_{0}^{s_{k - 1}}ds_{k}\times\]
\[\hspace{10mm}\times\prod_{i = 1}^{p - k}t_{i}^{-1-a-c-b-2(p-2)\rho}(1 -
t_{i})^{b - 1}\prod_{i<j}^{p - k}(t_{i} - t_{j})^{2\rho}\times\]
\[\hspace{10mm}\times\prod_{i = 1}^{k}s_{i}^{a}(1 - s_{i})^{c}\prod_{i <
j}^{k}(s_{i} - s_{j})^{2\rho} = \]
\[\hspace{10mm}= \prod_{i = 1}^{p -
k}\frac{\Gamma(i\rho)}{\Gamma(\rho)}\prod_{i = 0}^{p - k - 1}\frac{\Gamma(- a -
c - b - 2(p - 2)\rho + i\rho) \Gamma(b + i\rho)}{\Gamma(- a - c - 2\rho(k - 1)
- i\rho)}\times\]
\begin{equation}
  \hspace{10mm}\times\prod_{i = 1}^{k}\frac{\Gamma( i\rho )}{(\rho)}\prod_{i =
0}^{k - 1}\frac{\Gamma(1 + a + i\rho)\Gamma(1 + c + i\rho)}{\Gamma(2 + a + c +
(k - 2 + i)\rho)}
\end{equation}
\vspace{15mm}
Here the notations $v_{i} = t_{i}$ , if $1\leq{i}\leq{p - k}$ and $v_{i} =
s_{i}$ , if ${p - k + 1}\leq{i}\leq{p}$ are found to be suitable . The same
representation holds for the conformal blocks I$_{k}^{(0,q)}$(z) if one changes
 $a , b , c$ to the stressed ones as well as p to q .

To construct monodromy invarient physical correlation functions from the
obtained conformal blocks , one must sew the holomorphic
 conformal blocks with the corresponding antiholomorphic ones , by a monodromy
invariant metric , summing over the indice $k$ .

Due to the chosen above basis I$_{k}^{(p , 0)}$(z) ( I$_{k}^{(0 , q)}$(z) ) as
well as to the fact that the ( $a , a^{*}$ ) part of the conformal blocks
doesn't effect its monodromy properties ( as far as $\bigtriangleup{(a)} = 1$
and
$\bigtriangleup{(a^{*}) = 0 )}$ , the diagonal metric needed is the one
constructed for the minimal models , i.e. the $\chi_{k}$ . It can be derived
for the cases I$_{k}^{(p,0)}$(z) and I$_{k}^{(0,q)}$(z) in the same way as
in$^{2,3}$ .
The rezult coincides with the one for the minimal models up to an overall $k$
independant factor and appears to have the following form :
\vspace{10mm}
\[\chi_{k}^{(p , 0)}(a , b , c ; \rho) = \prod_{i = 1}^{p - k}s( i\rho
)\prod_{i = 0}^{p - k - 1}\frac{s(- a - c - b - 2(p - 2)\rho + i\rho)s(b +
i\rho)}{s(- a - c - 2\rho(k - 1) - i\rho)} \times \]
\vspace{5mm}
\begin{equation}
  \hspace{10mm}\times\prod_{i = 1}^{k}s( i\rho )\prod_{i = 0}^{k - 1}\frac{s(1
+
a + i\rho) s(1 + c + i\rho)}{s(2 + a + c + (k - 2 + i)\rho)} .
\end{equation}

The same formula holds for the case of (0 , q) conformal blocks , changing $a ,
b , c$ to $a', b', c'$ and p to q in (3.38) .

Finally , using all the calculations above and restoring the antiholomorphic
parts of the theory , one obtaines the following expression for the four point
correlation function (3.11) with the certain choices made in (3.12) and (3.13):
\vspace{5mm}
\[\langle \Phi_{H.W.}^{(\lambda_{1} , \mu_{1})}(z_{1} ,
\bar{z_{1}})\Phi_{H.W.}^{(\lambda_{2} , \mu_{2})}(z_{2} ,
\bar{z_{2}})\Phi_{T_{z}^{3} , Y^{3}}^{(\lambda_{3} , \mu_{3})}(z_{3} ,
\bar{z_{3}})\tilde\Phi_{L.W.}^{(\lambda_{4} , \mu_{4})}(z_{4} , \bar{z}_{4})
\rangle =   \]
\begin{equation}
  \hspace{10mm}\mid{z}\mid^{2\alpha_{12}}\mid{1 - z}\mid^{2\alpha_{23}}\prod_{s
< t}\mid{z_{s} - z_{t}}\mid^{-2\alpha_{st}} G(z , \bar{z} ).
\end{equation}
Here
\begin{equation}
   \alpha_{st} = \vec\beta^{(\lambda_{s} , \mu_{s})} C \vec\beta^{(\lambda_{t}
, \mu_{t})} ;
\end{equation}
\vspace{10mm}
\[G(z , \bar{z}) =
\lim_{R\rightarrow\infty}\mid{R}\mid^{4\tilde\bigtriangleup_{4}} \langle
\Phi_{H.W.}^{(\lambda_{1} , \mu_{1})}(0 , 0)\Phi_{H.W.}^{(\lambda_{2} ,
\mu_{2})}(z , \bar{z}) \Phi_{T_{z}^{3} , Y^{3}}^{(\lambda_{3} , \mu_{3})}(1 ,
1)
\tilde\Phi_{L.W.}^{(\lambda_{4} , \mu_{4})}(R , R) \rangle = \]
\vspace{5mm}
\begin{equation}
\hspace{5mm} = \sum_{k} S_{k}\mid{F_{k}}\mid^{2} .
\end{equation}
and
\begin{equation}
 S_{k} = \chi_{k}(N_{k})^{2} .
\end{equation}

An analogous four point correlation function can be biult easily from the
conformal blocks (3.16) .
\section{Concluding remarks}

The free filed representation was reviewed and some four point correlation
functions were constructed in this paper on the SU(3) conformal field theory .
The four point correlators considered were the ones with conformal blocks
having only
one type fo screening currents inserted . The latter were shown to have
Dotsenko - Fateev type structures$^{2,3}$ and perform the SU(2) subtheories of
the SU(3) conformal field theory .

Anyhow , it must be noted as well that there are indeed other conformal blocks
in the SU(3) conformal field theory containing the both $J_{1}^{+}$ and
$J_{2}^{+}$ currents and nevertheless reducible to the Dotsenko - Fateev type
integrals with
normalization integrals different in that they are multiplied by a polynomial
of a , b , c , a', b', c', and $\rho$ . But the consideration of such conformal
blocks (i.e. reducible ones) does not provide a complete basis for calculating
the structure
constants of the complete operator algebra in the way , performed in$^{4,15}$.

Moreover , it seems most likely that to create such a basis one must be able to
make use of the third screening current $J_{3}^{+}$ ( see e.g.$^{14,16,17}$ )
without which the construction of an arbitrary representation of the
$SL_{q}$(3) would be
impossible . Such a possibility , in its turn , could provide , in a certain
sense , an evidence of completness of the operator algebra structure constants
{}.

The usage of the third screening current $J_{3}^{+}$ in constructing conformal
blocks still needs to be investigated .

\vspace{15mm}
{\bf Acknowledgements}. I am grateful to Vl.S. Dotsenko for his stimulating
interest during the whole course of the work and for several fruitful
discussions . I have benefited much from the discussions with B.L. Feigin ,
A.A. Belavin , S.E. Parkhomenko ,
M.Y. Lashkevich and V.A. Sadov whom I express my gratitude as well.

\vspace{15mm}

\pagebreak
\vspace{10mm}
{\bf Figure captions}
\vspace{10mm}

{\bf Fig.1.}: The root diagram of SL(3) .

\vspace{5mm}

{\bf Fig.2.}: The weight diagram of the ($\lambda , 0$) representations .
\vspace{5mm}

{\bf Fig.3.}: The weight diagram of the ($0 , \mu$) representation .
\vspace{5mm}

{\bf Fig.4.}: The (1,0) , (2,0) and (3,0) representations .
\vspace{5mm}

{\bf Fig.5.}; The (0,1) , (0,2) and (0,3) representations .


\begin{thebibliography}{99}
\bibitem{1}A.A. Belavin , A.M. Polyakov and A.B. Zamolodchikov , Nucl. Phys.
B241 , 333 , (1984).
\bibitem{2}Vl.S. Dotsenko and V.A. Fateev , Nucl. Phys. B240 [FS12] , 312 ,
(1984).
\bibitem{3}Vl.S. Dotsenko and V.A. Fateev , Nucl. Phys. B251 [FS13] , 691 ,
(1985).
\bibitem{4}Vl.S. Dotsenko and V.A. Fateev , Phys. Lett. 154B , 291 , (1985).
\bibitem{5}B.L. Feigin and D.B. Fuchs , Moscow preprint , (1983).
\bibitem{6}M. Wakimoto , Commun. Math. Phys. 104 , 604 , (1986).
\bibitem{7}A.B. Zamolodchikov , unpublished.
\bibitem{8}V.A. Fateev and A.B.Zamolodchikov , Yad. Fiz. 43 , 1031 , (1986) ,
Sov. J. Nucl. Phys. 43 , 637 , (1986).
\bibitem{9}G. Felder , Nucl. Phys. B317 , 215 , (1986).
\bibitem{10}D. Bernard and G. Felder , ETH preprint , ETH - TH/89 - 26 ,
(1989).
\bibitem{11}B.L. Feigin and E.V. Frenkel , "Representations of affine Kac -
Moody algebras and bosonization" , in Collected papers to the memory of V.G.
Knizhnik , ( World Scientific ) .
\bibitem{12}V.G. Knizhnik and A.B. Zamolodchikov , Nucl. Phys. B247 , 83 ,
(1984).
\bibitem{13}Vl.S. Dotsenko , Nordita preprint , 89/54p , (1989).
\bibitem{14}P. Bouwknegt , J. McCarthy and K. Pilch , Commun. Math. Phys. 131 ,
125 , (1990).
\bibitem{15}Vl.S. Dotsenko , Santa Barbara preprint , NSF - ITP - 90 - 148.
\bibitem{16}P. Bouwknegt , J. McCarthy and K. Pilch , Phys. Lett. 234B , 297 ,
(1990).
\bibitem{17}P. Bouwknegt , J. McCarthy and K. Pilch , Progr. Theor. Phys.
Supll. 102 , 67 , (1990).
\end{thebibliography}
\end{document}